\documentclass[]{spie}  

\usepackage{amsfonts}
\usepackage{amsmath}

\usepackage{amsmath,amsfonts,amssymb}
\usepackage{graphicx}
\usepackage[colorlinks=true, allcolors=blue]{hyperref}

\title{Speech Motion Anomaly Detection via Cross-Modal Translation of 4D Motion Fields from Tagged MRI}

\author[a]{Xiaofeng Liu}
\author[a]{Fangxu Xing}
\author[b]{Jiachen Zhuo}
\author[c]{Maureen Stone}
\author[d]{Jerry L. Prince}
\author[e]{Georges El Fakhri}
\author[a]{Jonghye Woo}

\affil[a]{{Gordon Center for Medical Imaging, Massachusetts
General Hospital and Harvard Medical School, Boston, MA 02114 USA}}
\affil[b]{{Dept. of Radiology, University of Maryland School of Medicine, Baltimore, MD 21201 USA}}
\affil[c]{{Dept. of Neural and Pain Sciences, University of Maryland School of Dentistry, Baltimore, MD 21201 USA}}
\affil[d]{Dept. of Electrical and Computer Engineering, Johns Hopkins University, Baltimore, MD 21218 USA}
\affil[e]{Dept. of Radiology and Biomedical Imaging, Yale University, New Heaven, CT 06519, USA}

\begin{document} 
\maketitle

\begin{abstract}

Understanding the relationship between tongue motion patterns during speech and their resulting speech acoustic outcomes---i.e., articulatory-acoustic relation---is of great importance in assessing speech quality and developing innovative treatment and rehabilitative strategies. This is especially important when evaluating and detecting abnormal articulatory features in patients with speech-related disorders. In this work, we aim to develop a framework for detecting speech motion anomalies in conjunction with their corresponding speech acoustics. This is achieved through the use of a deep cross-modal translator trained on data from healthy individuals only, which bridges the gap between 4D motion fields obtained from tagged MRI and 2D spectrograms derived from speech acoustic data. The trained translator is used as an anomaly detector, by measuring the spectrogram reconstruction quality on healthy individuals or patients. In particular, the cross-modal translator is likely to yield limited generalization capabilities on patient data, which includes unseen out-of-distribution patterns and demonstrates subpar performance, when compared with healthy individuals.~A one-class SVM is then used to distinguish the spectrograms of healthy individuals from those of patients. To validate our framework, we collected a total of 39 paired tagged MRI and speech waveforms, consisting of data from 36 healthy individuals and 3 tongue cancer patients. We used both 3D convolutional and transformer-based deep translation models, training them on the healthy training set and then applying them to both the healthy and patient testing sets. Our framework demonstrates a capability to detect abnormal patient data, thereby illustrating its potential in enhancing the understanding of the articulatory-acoustic relation for both healthy individuals and patients.
  
\end{abstract}


\section{Introduction}

Accurate assessment of the association between tongue motion patterns during speech and the resulting speech acoustic outcomes is crucial in understanding how the movements of articulators, such as the tongue, affect the production of different speech sounds and their acoustic properties. This understanding is particularly critical, when evaluating and identifying abnormal articulatory features in patients with speech-related disorders, such as tongue cancer. It plays a pivotal role in accurately assessing speech quality and developing innovative treatment and rehabilitative strategies.

This work aims to develop a framework for detecting speech motion anomalies in conjunction with their corresponding speech acoustics. Anomaly detection~\cite{samariya2023comprehensive,che2021deep} is a fundamental problem in data mining. It is noted that a substantial portion of prior work in the domain of anomaly detection has primarily directed its attention towards datasets originating from a singular data source. In a multi-modal context, specific data instances frequently do not exhibit anomalies when examined independently within each distinct modality. Nonetheless, these instances may display incongruent patterns when multiple sources are collectively analyzed. In this work, the co-acquisition of tagged MRI and speech audio waveforms has been carried out to enable capturing both the movements of internal tongue motion during speech and the resulting speech acoustic outcomes. In particular, tagged MRI allows us to capture fine grained pixel-level tracking information of tongue motion during speech. The co-acquisition of multiple sources of information will facilitate the study to learn the relationship between tongue and oropharyngeal muscle deformations and their corresponding acoustic signals.

The joint analysis of disparate and heterogenous information, including tracking data and acoustic data, however, poses a significant challenge. The integration of information from diverse modalities has the potential to mutually enhance and improve detection performance. However, effectively handling complex distributions across these modalities necessitates the development of a principled framework for characterizing their inherent and complex correlations. Conventional linear models often struggle to capture such complex relationships effectively. To address this challenge, our prior work \cite{liu2022cmri2spec,liu2022taggedarxiv,liu2022taggedicassp,liu2023synthesizing,liu2023speech} has shown that 2D spectrograms converted from acoustic audio data are an effective proxy for this task, since 2D Mel-spectrograms have proven effective in capturing the distribution of acoustic energy across frequencies over time \cite{akbari2018lip2audspec,ephrat2017vid2speech,he2020image2audio}. 

Leveraging this representation, in this work, we present a novel framework that explicitly learns the healthy distribution using a deep cross-modal translator. This translator, which we previously developed, is trained with paired 4D motion fields and 2D spectrograms. Subsequently, our framework classifies new, unseen individuals as either normal or abnormal during testing. After training, the trained model is expected to perform well on healthy test data, accurately reconstructing spectrograms to produce intelligible audio. In contrast, patient data may contain inherent anomaly speech patterns that do not exist in the healthy training data. For the translator, we use either a well-performed fully 3D convolutional neural network (CNN) or 3D CNN+Longformer model for spatial temporal modeling \cite{liu2023synthesizing}. The prediction is then passed through the one-class SVM \cite{said2020network} for patient subject detection. Our results demonstrate that healthy and patient samples are distinguishable using our framework. Our framework marks the first effort to learn the mapping between 4D motion fields and audio waveforms using patient data for the purpose of anomaly detection. In a dataset collected from 39 subjects, our framework shows its capability to detect abnormal patient data, thereby demonstrating its potential in enhancing the understanding of the articulatory-acoustic relation for both healthy individuals and patients.

\begin{figure}[t]
\begin{center}
\includegraphics[width=1\linewidth]{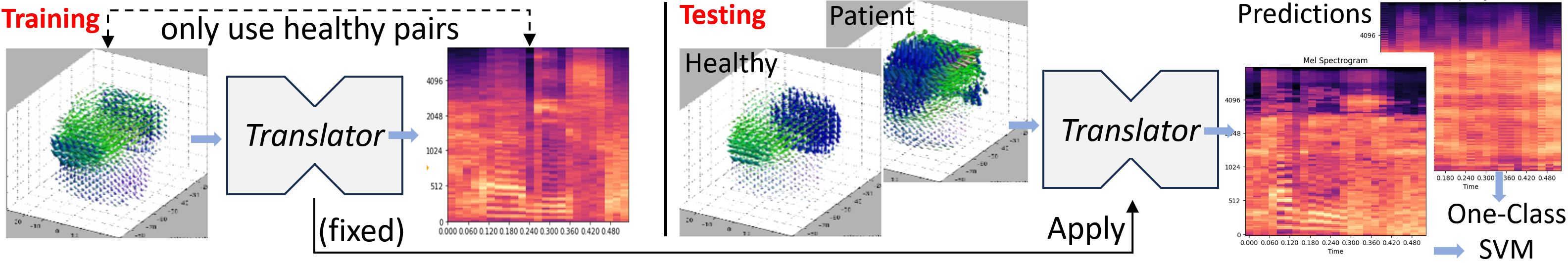}
\end{center}  
\caption{Illustration of our framework for unsupervised speech motion anomaly detection} 
\label{ccc}\end{figure}

\section{METHODS}\vspace{+5pt}

In this work, we acquired paired tagged MRI sequences and audio recordings from a total of 39 participants. The MRI data were acquired using a Siemens 3.0T TIM Trio system equipped with a 12-channel head coil and a 4-channel neck coil, using a segmented gradient echo sequence~\cite{lee2013semi,xing20133d}. Specifically, the dataset consists of a total of 20 individuals, with 19 healthy individuals and one patient, speaking the phrase ``a geese'', and another 19 individuals, with 17 healthy individuals and two patients, speaking the phrase ``a souk'' while following a periodic metronome-like sound.

We generated a sequence of voxel-level motion fields during the speech tasks from tagged MRI~\cite{xing2017phase,woo2021deep}. The resulting 4D motion fields $x$ consist of 26 frames with a size of ${3\times[(128\times128\times128)\times26]}$, where each voxel of the motion fields has three channels to represent 3D directions. As for the corresponding audio waveforms, their lengths range from 21,832 to 24,175. To augment the data, we employed a sliding window technique to crop the audio waveforms to a fixed length of 21,000, resulting in 100$\times$ audio waveforms per recording. Subsequently, we utilized the publicly available Librosa library\footnote{\href{https://librosa.org/doc/main/generated/librosa.feature.melspectrogram.html.}{Librosa: generating mel-spectrogram from audio waveforms.}} for further processing. The audio waveforms were converted into Mel-spectrograms with a size of $64\times64$. Of note, the Mel-spectrogram utilizes the mel-scale, which is a non-linear transformation of the Hz-scale, to accentuate frequencies within the human voice range spanning from 40 to 1000 Hz, while attenuating high-frequency instrument noise.

We trained a translator $\mathcal{T}$ from 4D motion fields to their corresponding 2D mel spectrograms using healthy individuals only. We used two different backbones that we previously developed, demonstrating superior performance in this task~\cite{liu2023synthesizing}. The fully 3D CNN model uses 3D CNN for both spatial and temporal modeling, while the 3D CNN+Longformer model uses the Longformer framework to efficiently model temporal correlation~\cite{liu2023synthesizing}. In addition, the generative adversarial networks (GAN) is further added to improve the realism as previous work~\cite{liu2023synthesizing}. However, previous studies \cite{liu2023synthesizing} did not differentiate between healthy and patient data, making them not able to support speech motion anomaly analysis. Furthermore, incorporating any speech anomaly cases into the study is challenging, as it requires learning solely from healthy individuals and efficiently detecting outliers.

The model is expected to learn the translation on healthy data and demonstrate good performance on healthy testing individuals. By contrast, the data from patients can be considered as outlier samples, as they may exhibit anomalous speech motion patterns. In particular, deep learning is highly dependent on the assumption of i.i.d. of training and testing data, and cannot generalize well on unseen, different data~\cite{liu2022deep}. Thus, we can distinguish between healthy and patient data, based on the performance of our translator $\mathcal{T}$, which was trained solely on healthy data.

It is important to note that directly applying a threshold from the reconstruction error for classification may not yield satisfactory results due to the relatively large variance observed during cross-validation~\cite{said2020network}. Thus, we adopted the one-class SVM with non-linear RBF kernal\footnote{\href{https://scikit-learn.org/stable/auto_examples/svm/plot_oneclass.html}{One-class SVM with non-linear kernel (RBF)}} trained on healthy data to classify the unseen patient data as similar or different to the healthy training set. It is worth mentioning that the one-class SVM achieves unsupervised novelty detection, by learning the boundaries of normal points to classify any points lying outside the boundary as outliers~\cite{scholkopf2000support}. The training and testing procedures are illustrated in Fig. 1.

\begin{table}[t]
\centering 
\caption{Numerical comparisons of different methods for testing on healthy or patient data. $\uparrow$ the higher, the better.} \vspace{+8pt}
\resizebox{0.85\linewidth}{!}{
\begin{tabular}{l|c|c|c}
\hline
Backbone Model&  Testing Subjects  & ~~~Corr2D for spectrogram $\uparrow$~~~ & ~~~PESQ for waveform $\uparrow$~~~\\\hline\hline


3D CNN + Transformer~\cite{liu2023synthesizing}& Healthy & \textbf{0.814}$\pm$0.017  &  \textbf{1.638}$\pm$0.019 \\ 

  & Patient	&  {0.786}$\pm$0.015  &   1.611$\pm$0.022 \\\hline \hline

Two-stage 3D CNN~\cite{liu2023synthesizing}  &  Healthy 	   &\textbf{0.805}$\pm$0.018  &  \textbf{1.624}$\pm$0.020\\ 

& Patient	   &{0.773}$\pm$0.021  &  1.604$\pm$0.017\\\hline
\end{tabular}} 
\label{tabel:1}  
\end{table}

\section{RESULTS}

\begin{figure}[t]
\begin{center}\vspace{+15pt}
\includegraphics[width=1\linewidth]{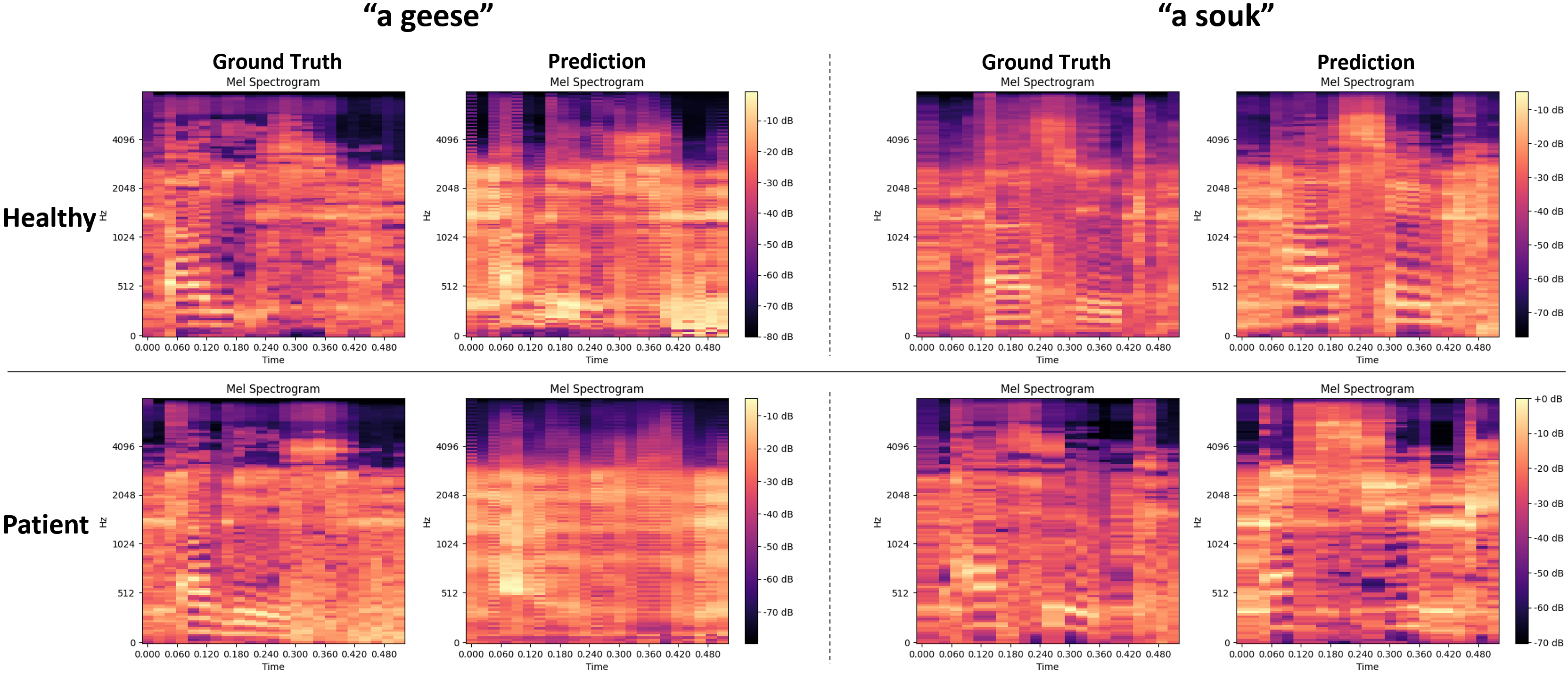}
\end{center}  \vspace{-5pt}
\caption{An example of the synthesized spectrograms from healthy and patient motion fields using 3D CNN+Longformer} 
\label{ccc}\end{figure}

We adopted the same translator backbones and hyperparameters as proposed in our previous work~\cite{liu2023synthesizing}. We implemented our framework using PyTorch and conducted training on an NVIDIA V100 GPU for a total of 200 epochs. With the translation model trained on healthy individuals, we compared its performance on healthy or patient testing sets. In the case of healthy individuals, we performed a subject-independent leave-one-out evaluation, while retaining all patient data for testing. In each round, we used a total of 35 healthy individuals for training, and test on one healthy individual and three patients.

 
In order to evaluate the fidelity of the generated spectrogram in the frequency domain and the audio waveform in the time domain, we employed the 2D Pearson's correlation coefficient (Corr2D) and the Perceptual Evaluation of Speech Quality (PESQ), respectively, following previous studies~\cite{akbari2018lip2audspec,liu2022tagged}. Elevated values of Corr2D and PESQ signify superior synthesis performance. The averaged results from all 36 rounds are presented in Table 1.

Notably, both Corr2D and PESQ metrics show that the performance on patients is inferior compared to that on healthy individuals. An example of the predicted spectrogram and audio waveform is shown in Fig. 1. This example illustrates that audio from healthy individuals can be synthesized from a sequence of motion fields, whereas there is notably greater inconsistency observed in the case of patient data.
 
\vspace{-5pt}
In Fig. 2, we present an Receiver Operating Characteristic (ROC) curve of the one-class SVM, which incorporates cross-validation variability. The area under the curve (AUC) significantly exceeds the chance level AUC of 0.5, suggesting that the spectrograms generated from healthy individuals and patients are effectively distinguishable.

\begin{figure}[t]
\begin{center}\vspace{+10pt}
\includegraphics[width=0.8\linewidth]{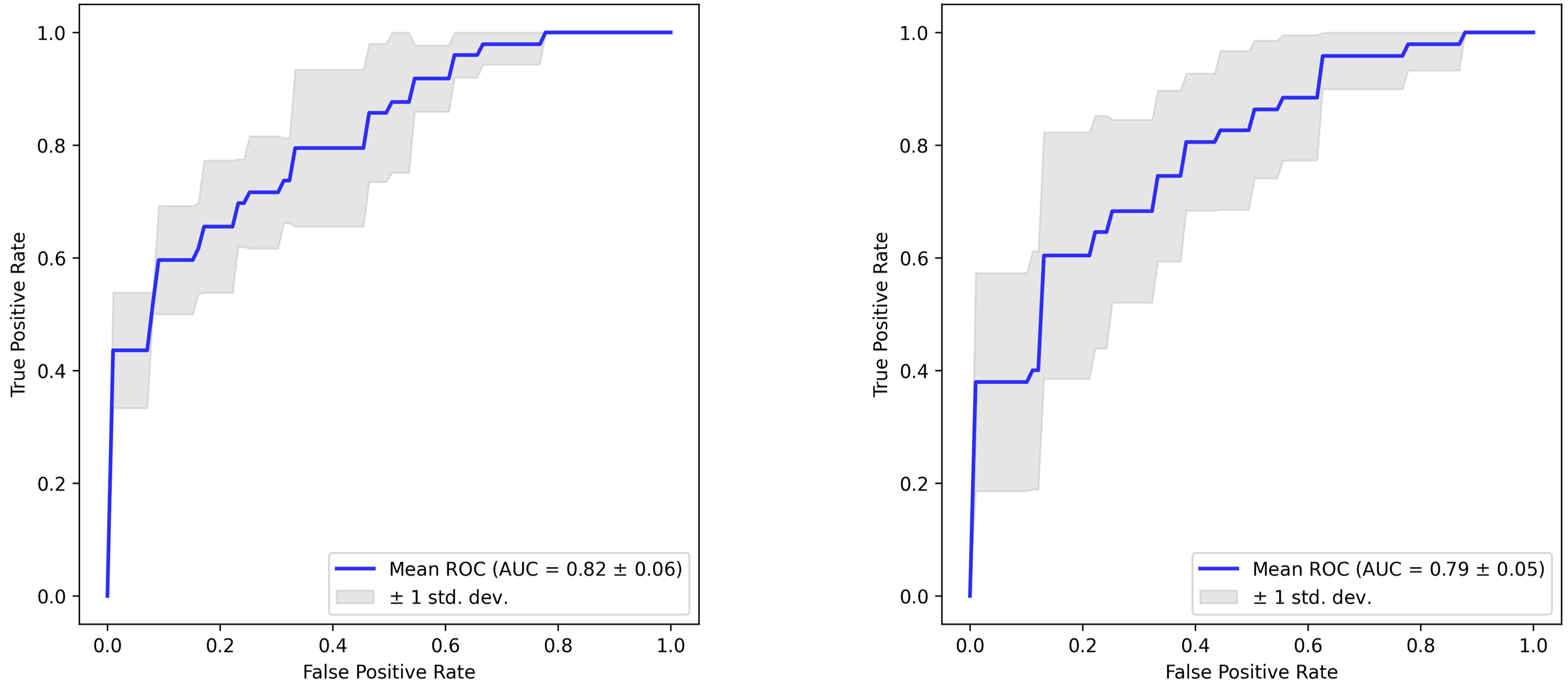}
\end{center}  
\caption{An ROC curve for the classification of healthy and patient groups with cross-validation variability using 3D CNN+Transformer (left) and 3D CNN (right) with the one-class SVM.} 
\label{ccc}\end{figure}




 \vspace{-5pt}
\section{CONCLUSION}

In this work, we proposed an unsupervised speech motion anomaly detection framework that exclusively learns from healthy individuals. After learning from the healthy sequence of motion fields to the corresponding spectrograms, our framework demonstrates varying performance between healthy and patient groups. Our framework makes use of the one-class SVM to effectively differentiate between the two groups. Our framework can be easily generalized to different speech-related disorders, facilitating a better understanding of the correlations between tongue motion and the resulting speech acoustic outcomes. In future work, we plan to carry out a more detailed analysis of the correlations between patient MR data and the reconstructed audio.


\vspace{+5pt}
\acknowledgments 

This work is supported by NIH R01DC014717, R01DC018511, and R01CA133015.


\bibliography{main} 
\bibliographystyle{spiebib} 

\end{document}